\documentclass[preprint,12pt]{article}
 
\usepackage{graphicx} 
\usepackage{caption}
\usepackage{subcaption}

\usepackage{amssymb}
\usepackage{amsthm}
\usepackage{amsmath}
\usepackage{amsfonts}
\usepackage{mathptmx}
\usepackage{mathtools}

\usepackage{bm} 
\usepackage{authblk}

\usepackage{algorithm}
\usepackage{algpseudocode} 

\title{TTCF4LAMMPS: A toolkit for simulation of the non-equilibrium behaviour of molecular fluids at experimentally accessible shear rates}

\author[1]{Luca Maffioli \thanks{\texttt{lmaffioli@swin.edu.au}}}
\author[2]{James P. Ewen}
\author[3]{Edward R. Smith}
\author[1,5]{Sleeba Varghese}
\author[4]{Peter J. Daivis}
\author[2]{Daniele Dini}
\author[1]{B. D. Todd}

\affil[1]{Department of Mathematics, School of Science, Computing and Engineering Technologies, Swinburne University of Technology, P.O. Box 218, Hawthorn, 3122, Victoria, Australia}

\affil[2]{Department of Mechanical Engineering, Imperial College London, South Kensington Campus, London, SW7 2AZ, London, United Kingdom}

\affil[3]{Mechanical and Aerospace Engineering, Brunel University London, Kingston Lane, Uxbridge, UB8 3PH, London, United Kingdom}

\affil[4]{School of Science, RMIT University, P.O. Box 2476, Melbourne, 3001, Victoria, Australia} 

\affil[5]{``Glass and Time'', IMFUFA, Department of Science and Environment, Roskilde University, Roskilde, 4000P.O. Box, Denmark}

\begin{document}
	\maketitle

\begin{abstract}
We present TTCF4LAMMPS, a toolkit for performing non-equilibrium molecular dynamics (NEMD) simulations to study fluid behaviour at low shear rates using the LAMMPS software. By combining direct NEMD simulations and the transient-time correlation function (TTCF) technique, we study the behaviour of fluids over shear rates spanning $15$ orders of magnitude. We present two example systems consisting of simple monatomic systems: one containing a bulk liquid and another with a liquid layer confined between two solid walls. The small bulk system is suitable for testing on personal computers, while the larger confined system requires high-performance computing (HPC) resources. We demonstrate that the TTCF formalism can successfully detect the system response for arbitrarily weak external fields. We provide a brief mathematical explanation for this feature. Although we showcase the method for simple monatomic systems, TTCF can be readily extended to study more complex molecular fluids. Moreover, in addition to shear flows, the method can be extended to investigate elongational or mixed flows as well as thermal or electric fields. The reasonably high computational cost needed for the method is offset by the two following benefits: i) the cost is independent of the magnitude of the external field, and ii) the simulations can be made highly efficient on HPC architectures by exploiting the parallel design of LAMMPS. We expect the toolkit to be useful for computational researchers striving to study the nonequilibrium behaviour of fluids under experimentally-accessible conditions.
\end{abstract}

Over the last few decades~\cite{Evans1986}, nonequilibrium molecular dynamics (NEMD) simulations have provided atomic-scale insights into important fluid behaviour under shear such as turbulence~\cite{Smith2015}, cavitation~\cite{Savio2016}, boundary slip~\cite{Martini2008}, and shear thinning~\cite{Bair2002,Jadhao2017}.
One significant limitation of these NEMD simulations is that high shear rates are required to obtain satisfactory signal-to-noise ratios~\cite{Todd2017,Ewen2018}.
In general, the shear rates required are higher than those that can be applied experimentally by several orders of magnitude, which prevents direct experimental validation of the simulations~\cite{Bair2002,Jadhao2017}.
This problem can be overcome using the transient-time correlation function (TTCF) formalism~\cite{evans1990statistical,evans2016fundamentals}, which is a non-linear generalization of the Green-Kubo formula~\cite{Green1954,Kubo1957}.
TTCF utilises the time correlation between the so-called dissipation function of the system and the transient response of any arbitrary phase variable after an external field is activated.
The method forms a bridge between equilibrium techniques to calculate transport coefficients (e.g. shear viscosity, diffusion constant, and thermal conductivity) where no external field is applied and direct NEMD where strong fields are required~\cite{Sarman1998}.
TTCF has been successfully applied to study the rheology of bulk monatomic fluids~\cite{Morriss1987,Evans1988,Borzsak2002,Desgranges2009}, molecular fluids~\cite{Pan2006,Mazyar2009}, and liquid metals~\cite{Desgranges2008,Desgranges2008a} at low shear rates.
In addition to shear flows, TTCF has been used to study elongational flows~\cite{Todd1997} and mixed flows~\cite{Hartkamp2012}.
It has also been applied to investigate monatomic~\cite{Delhommelle2005,Bernardi2012,Maffioli2022} and molecular fluids~\cite{Bernardi2016} confined between sliding solid surfaces.
TTCF can also be used to investigate other types of external fields such as electric fields~\cite{Delhommelle2005a,English2010,English2011} of strengths that are closer to those applied experimentally compared to direct NEMD simulations.
Evans et al.~\cite{evans2016fundamentals} derived the theoretical framework for a wider range of nonequilibrium phenomena including heat transfer in inhomogeneous systems and the relaxation from a nonequilibrium thermodynamic state.

Despite their clear benefits over direct NEMD simulations, TTCF-based NEMD simulations have only been utilised by a handful of research groups worldwide, arguably due to the complexity of the method implementation.
Here, we present TTCF4LAMMPS, a toolkit for studying the non-equilibrium behaviour of fluids under weak external fields using the open source Large-scale Atomic/Molecular Massively Parallel Simulator (LAMMPS) software~\cite{thompson2022lammps}.
The method can be made very efficient on HPC architectures by utilising the highly parallel nature.
We demonstrate the application of the TTCF method to two example systems consisting of simple monatomic fluids under shear: one bulk and one confined.

\section*{Mathematical Framework}

Traditional NEMD simulations require the user to monitor the system over a single nonequilibrium trajectory from which the response is measured. A time average is then performed over the data sampled from the steady state. The process can be optimized by running in parallel multiple independent copies of the same system, where each system starts from a different initial condition. Each system is in the same thermodynamic state (temperature, density, etc), is subjected to the same force and hence undergoes analogous transient toward the nonequilibrium steady state. The time average is then replaced by an ensemble average, where the signal at time $t$ is the average of the response across all the systems at time $t$. 
The Transient Time Correlation Function (TTCF) technique is instead based on the following result of nonequilibrium statistical mechanics\cite{evans1990statistical}:
\begin{equation}
	\label{eqn:TTCF}
	\langle B(t)\rangle=\langle B(0)\rangle+\int_0^t\langle\Omega(0)B(s)\rangle\text{d}s
\end{equation}
where $B(t)$ is the arbitrary quantity measured from the system. The formula states an identity between the phase space average (left hand side) of $B(t)$ and the integral of the time correlation between the same variable $B(t)$ and the so-called dissipation function $\Omega$ (right hand side). $t=0$ represents the beginning of the nonequilibrium trajectory, when the external field is switched on and the system is driven out of equilibrium.
$\Omega$ is equal to $\beta\dot{H}^{ad}$, with $\beta=1/k_BT$ and $\dot{H}^{ad}$ the adiabatic time derivative of the internal energy, that is, the derivative of the mechanical energy without accounting for any thermostat term. $k_B$ and $T$ are the Boltzmann constant and the temperature set by a thermostat.  
For the application of TTCF, two conditions must be met: first, the initial conditions for the nonequilibrium trajectories must be generated by the equilibrium probability distribution of the system.
In practice, one must follow the system over an equilibrium $\textit{mother}$ trajectory, from which the microscopic state $\bm{\Gamma}(t_i)$ is periodically sampled.  $\bm{\Gamma}(t_i)$ is the collection of all positions and momenta of the particles at time $t_i$, and is used as the initial condition for the $i$-th nonequilibrium run, or $\textit{daughter}$ trajectory.
To get good statistics, many initial states and corresponding daughter trajectories must be produced. The mother trajectory is hence a simple tool to generate the correct ensemble over which to compute the phase space average. In Eq. \eqref{eqn:TTCF},  $t=0$ corresponds to the initial state of the nonequilibrium run, when the system is still at equilibrium and the external force is activated. The response $B(t)$ is then monitored over each daughter trajectory, correlated with the dissipation function $\Omega$ at $t=0$, the average is performed at each time step across all the trajectories, and integrated as per Eq. \eqref{eqn:TTCF}. The resulting integrand function is then
\begin{equation}
	\label{eqn:integrand}
	\langle\Omega(0)B(s)\rangle=\dfrac{1}{N}\sum_i ^N \Omega(0)_iB(s)_i
\end{equation}
where $N$ is the number of daughter trajectories generated, and $\Omega(0)_i$, $B(s)_i$ are the dissipation function and the response computed from the $i$-th trajectory.

The second requirement is that the system is \textit{mixing}, that is, $\Omega(0)$ and $B(t)$ must eventually decorrelate, $\langle\Omega(0)B(t)\rangle \rightarrow \langle\Omega(0)\rangle\langle B(t)\rangle$  for $t \to \infty$. Since at $t=0$ the system is at equilibrium, $\langle \Omega(0)\rangle=0$, and the convergence of the integral is ensured. The mixing requirement prevents the direct application of the TTCF algorithm to systems characterized by long time correlations, which is common, for instance, in the solid state. 
Since $\Omega(0)$ is null only on average ($\langle\Omega(0)\rangle=0$ but in general $\Omega(0)_i \ne 0$), a finite sample can hardly guarantee $\langle\Omega(0)\rangle=0$ and hence the perfect convergence of the integral. The issue can be avoided by generating further initial conditions from each state $\bm{\Gamma}_i$ by using the following transformations:
\begin{equation}
	\label{mappings}
	\begin{split}
		\bm{\Gamma}_i=\bigl(\textbf{x},\textbf{y},\textbf{z},\textbf{p}_x,\textbf{p}_y,\textbf{p}_z\bigr)&\longrightarrow\bm{\Gamma}'_i=\bigl(\textbf{x},\textbf{y},\textbf{z},-\textbf{p}_x,-\textbf{p}_y,-\textbf{p}_z\bigr)\\
		\bm{\Gamma}_i=\bigl(\textbf{x},\textbf{y},\textbf{z},\textbf{p}_x,\textbf{p}_y,\textbf{p}_z\bigr)&\longrightarrow\bm{\Gamma}''_i=\bigl(-\textbf{x},\textbf{y},\textbf{z},-\textbf{p}_x,\textbf{p}_y,\textbf{p}_z\bigr)\\
		\bm{\Gamma}_i=\bigl(\textbf{x},\textbf{y},\textbf{z},\textbf{p}_x,\textbf{p}_y,\textbf{p}_z\bigr)&\longrightarrow\bm{\Gamma}'''_i=\bigl(-\textbf{x},\textbf{y},\textbf{z},\textbf{p}_x,-\textbf{p}_y,-\textbf{p}_z\bigr)\\
	\end{split}
\end{equation}
where the first mapping inverts the sign of the momentum for the particles. 
The second mapping is an $x$-reflection of the entire system, while the third is a combination of the first and the second. For a Couette flow applied in the $xy$ plane, $\bm{\Gamma}''$ and $\bm{\Gamma}'''$ change the sign of $\Omega(0)$ and hence the average over the four states (original + mappings) is identically null. It is important to note that these mappings, particularly the second and the third, are effective \textit{only} if the system is subjected to a shear in the $xy$-plane. Different types of external forces may require other mappings.
There are several different choices of transformations: each of the position and momentum component can be mirrored independently, and some authors even employed a permutation of the components \cite{Desgranges2009} (if the system is periodic and has the same size along the $x$ and $z$ dimension, $\textbf{x}$ and $\textbf{z}$ and the corresponding momenta can be swapped), but any combination has to meet the condition $\langle \Omega(0)\rangle=0$, and each mapping must have the same probability of the original state of being sampled from the mother trajectory. Alternatively, the integrand function can be modified with
\begin{equation}
	\label{corr_modified}
	\langle\Omega(0)B(s)\rangle-\langle\Omega(0)\rangle\langle B(s)\rangle
\end{equation}
which automatically includes the correction for finite sample effects. This second expression is redundant if the mappings guarantees $\langle\Omega(0)=0\rangle$, but was successfully applied in a previous work \cite{Bernardi2012}, where the proper transformations were not used.

\begin{figure}
	\centering
	
	\includegraphics[width=\linewidth]{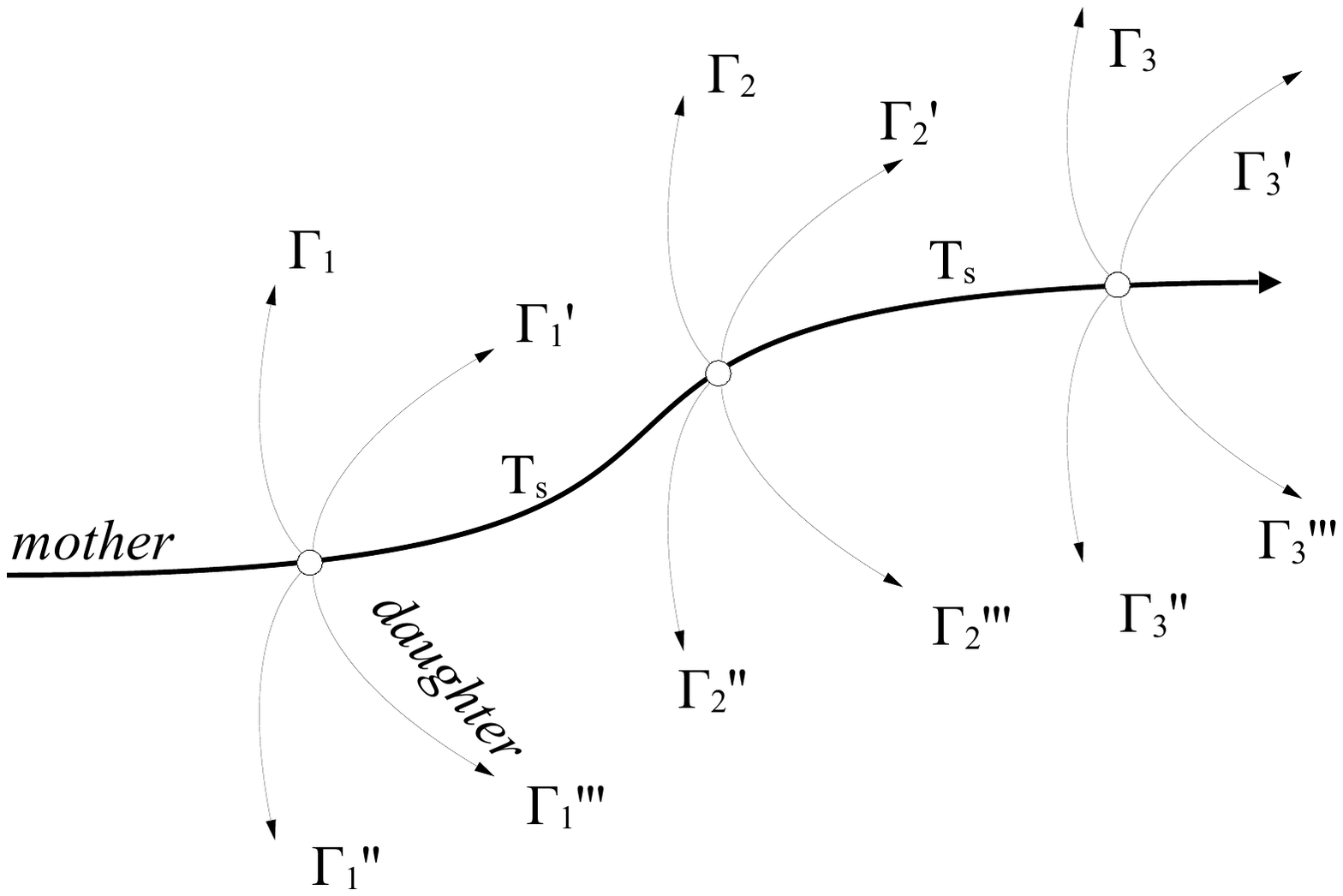}
	
	\caption{Schematic representation of the TTCF method. From the equilibrium mother trajectories, the microscopic state is sampled every $N_s$ time units. Four daughter nonequilibrium trajectories depart form the initial state.}
	\label{fig:TTCF}
\end{figure} 

The general scheme of the TTCF algorithm is shown in Figure \ref{fig:TTCF}. The system is followed along an equilibrium mother trajectory. After a period of thermalization, when the system has fully reached the thermodynamic equilibrium, the sampling process starts. Every $T_s$ time units the microstate $\bm{\Gamma}(t)$ is stored, and used as an initial condition for the corresponding daughter trajectories, after being properly modified with the mappings selected. It is important to clarify that the dissipation function should be computed on the daughter trajectory, after the external filed has been switched on. The delay $T_s$ between samples should be long enough to ensure an independent set of initial states. The autocorrelation function of some benchmark quantities can be used to determine the necessary delay. These variables can be the same function $B$ that will be monitored along the nonequilibrium segments, or $\Omega(t)B(t)$.
The class of applicability of the TTCF formalism is very broad. Since molecular bonds are conservative forces or ideal constraints, they do not bring any contribution to the dissipation function, which is instead completely determined by the external force driving the system out of equilibrium. Moreover, the implementation of a TTCF algorithm is feasible for several types of external field 
\cite{evans2016fundamentals}.
\section*{Software methodology}
Due to its computational cost, TTCF should primarily be implemented on HPC clusters. The number of daughter trajectories needed is typically in the order of tens of thousands or even millions. Hence, three major bottlenecks can be found which significantly impact the usage of the method on local machines: total simulation time, total memory size of the output, and number of output files produced. While custom built codes can be designed to minimize the size and number of output files, e.g. by including the averaging and integration process in the same script, this feature may sometimes be cumbersome to implement with molecular dynamics packages such as LAMMPS. However, a TTCF script can be run on local machines for particularly simple systems, assuming the outputs are carefully managed. This operation can be obtained by combining the internal structures and functions available in LAMMPS and its Python interface which allows a Python script to handle LAMMPS scripts. 

In this paper, we provide two different TTCF implementations. The first is designed to be run on HPC clusters. It is more flexible and intuitive, but it generates a large number of output files and requires a large amount of mass storage. The second paradigm is suitable also for local machines. Its LAMMPS script is designed to generate both mother and daughter trajectories in a single run. It does not produce any output files except for the final result. It is more compact and optimized, and can be faster if the simulation itself is particularly quick, and the process of reading and writing on file becomes relatively intensive. 

The first scheme works as follows: since the number of nonequilibrium runs required for the TTCF calculation is almost certainly much larger than the number of available cores, the total number of daughter trajectories is split into several independent, single-core runs. Each run is composed of an equilibrium mother trajectory which generates the required number of initial states, after which the corresponding nonequilibrium simulations are performed. Since each run is independent of the others, it can start as soon as a single core is available on the cluster. This scheme is the so-called ``embarrassingly parallel'' setup, and, as a result, the simulation is extremely fast to run, even if the overall amount of resources needed is large. The pseudocode of the corresponding workflow for a shell script is
\begin{algorithm}
	\caption{SHELL SCRIPT}\label{SHELL}
	\begin{algorithmic}[1]
		\For{$i = 1, \dots, \text{Ncores}$}
		\State select seed SEED($i$) from list of random integers
		\State run LAMMPS MOTHER ( SEED($i$) )
		\State	\textit{produce} Ndaughters \textit{initial states (files)}
		\For{$n = 1, \dots, \text{Ndaughters}$}
		\For{$m = 1, \dots, \text{Nmaps}$}
		\State run LAMMPS DAUGHTER ($n$,$m$)
		\State\textit{take initial state (file) n, apply mapping m}
		\State\textit{produce} $\text{OUTPUT}\textunderscore\text{FILE}_{n,m}$
		\EndFor
		\EndFor
		\State perform partial average and TTCF integration
		\EndFor
		\State gather total averages, total TTCF integration, statistics.
	\end{algorithmic}
\end{algorithm}
the script cycles over the required cores to launch independent runs, as depicted in Algorithm \ref{SHELL}. Each mother trajectory is initialized by a random number, accordingly select from a list previously created. The random number is the seed for the random velocities with which the particles are initialied at the start of the run. LAMMPS run then generates $Ndaughters$ files which contain the collection of the starting points for the daughter trajectories. The script then cycles over each state $n$ and each mapping $m$. The LAMMPS daughter script reads the file $n$ and applies the mapping $k$, and runs the nonequilibrium simulation from which the response is monitored, and stored in the file indexed as $n,m$. At the end of the double loop, the data within each core are averaged, and a first TTCF integration is performed over the partial results. Once the partial results have been produced, a final script cycles across each of them to perform the final average, the final TTCF integration, and to estimate the standard error of the results.
The process of writing and reading the initial states should be perfomed with the LAMMPS command \emph{write\textunderscore restart} and \emph{read\textunderscore restart}, which ensure the whole system's state is transferred from mother to daughter trajectory including the value of the thermostat, if relevant.

The second scheme is designed to maximize the efficiency. Hence, we manage and run the LAMMPS simulation via a Python script. The outputs are not written on file but directly managed by the Python code, which averages the data from all the daughter trajectories and stores them until the end of the simulation, when the output file is eventually produced. 

From a computational perspective, the only difference between mother and daughter trajectory is the action, on the latter, of the external field, and the computation of system's response. Hence, the same LAMMPS script can be employed to generate both mother and daughter trajectory, without the need to run the daughter trajectories via a different script. 

\begin{figure}
	\centering
	\includegraphics[width=1\linewidth]{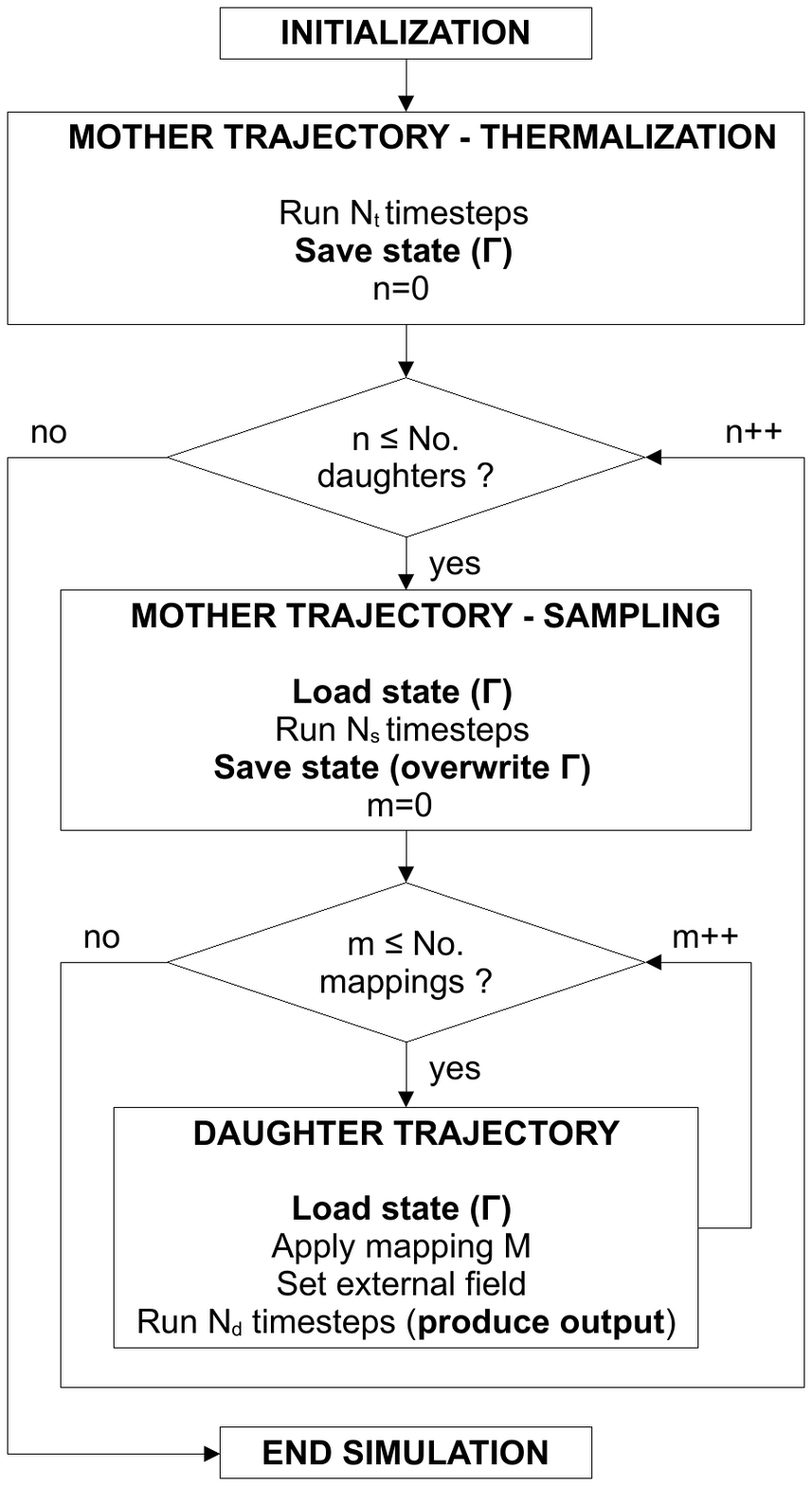}
	\caption{Flowchart of the LAMMPS script for TTCF calculation through a single uninterrupted run. Note that the sequence of commands \textbf{save state} and \textbf{load state} at the end of the thermalization and at the beginning of the outer cycles, as well as the first decorrelation (Run $N_s$ timesteps) are redundant, but necessary to preserve the correct structure of the flow.}
	\label{fig:flowchart}
\end{figure}

The procedure works as follows. After the thermalization, the sampling process starts. The system is followed over the equilibrium thermostatted dynamics (\emph{fix nvt} command) for the desired number of timesteps. 
The sample $\bm{\Gamma}_n$ is then produced by storing positions and momenta via the command \emph{fix store/state}. Once a sample is generated, the mother trajectory is halted (\emph{unfix nvt}), and the script cycles over the daughters. 
Each daughter run is characterized by the following sequence of operations: set the system to state $\bm{\Gamma}_n$ via \emph{set atom}, where the target values are the outputs of the \emph{fix store/state}; modify the initial state accordingly to the mapping; set the nonequilibrium dynamics, call the function for the computation of the desired quantities (\emph{fix ave/time}, \emph{fix ave/chunck}, etc), and run the simulation. 
At the end of each NEMD trajectory, the nonequilibrium dynamics is halted as well as the computation of the response (\emph{unfix ave/time}, \emph{unfix ave/chuck}, etc). The process is cycled over all the four daughters. 
After the nonequilibrium runs have been generated, the system is set back to the state $\bm{\Gamma}_n$ and the mother trajectory is recovered. The system is then followed over the equilibrium dynamics until the next sample $\bm{\Gamma}_{n+1}$ overwrites the previous state $\bm{\Gamma}_n$, when the cycle starts again. The loop is repeated the desired number of times. The scheme is shown in the Algorithm \ref{LOCAL}
\begin{algorithm}
	\caption{LAMMPS SCRIPT}\label{LOCAL}
	\begin{algorithmic}[1]
		\State initialize system (null ext. field)
		\State run $N_t$ timestep
		\State save state
		\For{$n = 1, \dots, \text{Ndaughters}$}
		\State load system state
		\State run $N_s$ timestep
		\State save system state
		\For{$m = 1, \dots, \text{Nmaps}$}
		\State load system state
		\State apply mapping
		\State switch on ext. field
		\State set computation of the response
		\State run $N_d$ timestep
		\State switch off ext. field
		\EndFor
		\EndFor
	\end{algorithmic}
\end{algorithm} 
and in Figure \ref{fig:flowchart}. The output of the functions \emph{fix ave/time}, \emph{fix~ave/chunck} are not written to file, but passed to the Python script through which the LAMMPS script is launched. The loss in efficiency caused by the Python wrapper is abundantly compensated by a easier management of the simulations, since only few files are generated, and by avoiding reading/writing on files, which is typically inefficient both on a local machine and on a HPC cluster \cite{HUANG2022106}
Since \emph{fix store/state} does not save the value of the thermostats, the periodically halted and resumed mother trajectory diverges from the one obtained via a single, uninterrupted run. The effect occurs at the transition from mother to daughter runs as well. The issue is irrelevant for the mother trajectory if the runs between samples are sufficiently long to fully regain an equilibrium canonical distribution. Moreover, it has an undetectable effect on the shear pressure, but it should potentially be accounted for if other quantities, such as the temperature, are monitored. In that case, the initial states should be stored and loaded via the commands \emph{write\textunderscore restart} and \emph{read\textunderscore restart}, and daughter and mother trajectories should have identical \emph{fix\textunderscore nvt} commands. However, not every \emph{fix} is stored in a \emph{restart} file, and hence the transition from mother to daughter trajectory might not always be totally consistent.

Since the thermalization is typically very long, both setups can be optimized by running a smaller number of independent thermalizations over multicore runs. For instance, instead of 1000 thermalizations over single core jobs, 100 thermalizations with 10-core independent jobs can be performed. After the process is completed, the workflow is switched back to single core jobs. Hence, for each original trajectory, 10 further mother trajectories must be generated, on which the sampling of the initial states take place. It is easy to generates further initial conditions from a single state by adding to the velocities a small random kick. Since the system is chaotic, the decorrelation between the trajectories is fast, and a small perturbation should not alter the thermodynamic state of equilibrium achieved through the thermalization. The precise gain in efficiency of this method is rather situational, since it depends, among other things, on the specific architecture of the cluster used for the simulation.   
\section*{Example 1: shear of bulk systems}

The bulk system is based on that used in a previous study by Borzsák et al. \cite{Borzsak2002}.
It consists of 256 particles placed in a cubic box of fixed volume.
Periodic boundary conditions are applied in all three Cartesian directions.
The evolution is driven by the SLLOD equations of motion \cite{Evans1984,Evans1986,evans1990statistical,Daivis2006,Todd2017}. In LAMMPS, the dynamics is described by the following command:
$fix~deform$ which deforms the simulation box according to the selected shear rate, in order to give the fluid a linear Couette flow profile.
This approach is essentially equivalent to the popular Lees-Edwards sliding-brick boundary conditions \cite{Lees1972}.
Finally \emph{fix nvt/sllod} produced a thermostatted SLLOD dynamics, where the temperature is controlled with a Nos{\'{e}}-Hoover thermostat \cite{Nose1984,Hoover1985}, and the equations integrated using the velocity-Verlet (or Störmer–Verlet) algorithm \cite{Verlet1967} (using the \emph{run\textunderscore style verlet} command).
The thermostatted SLLOD dynamics is described by the following set of differential equations
\begin{equation}
	\label{eqn:system1}
	\begin{split}
		\dot{\textbf{r}}_i&=\dfrac{\textbf{p}_i}{m_i}+\textbf{i}\dot{\gamma}y_i\\
		\dot{\textbf{p}}_i&=-\sum_j\nabla\phi_{ij}-\textbf{i}\dot{\gamma}p_{yi}-\alpha\textbf{p}_i\\
		\dot{\alpha}&=\dfrac{1}{Q}\biggl(\sum_i\textbf{p}^{2}_i-3Nk_BT\biggr)\\
	\end{split}
\end{equation} 
where $\alpha$ is the is the Nos{\'{e}}-Hoover thermostat  \cite{Nose1984,Hoover1985} and $Q$ its damping factor.
The particles interact via the Weeks-Chandler-Andersen (WCA) potential \cite{Weeks1971}, which is a truncated and shifted Lennard-Jones potential \cite{Jones1924}.
\begin{equation}
	\phi\left(r_{ij}\right)= 
	\begin{cases}
		4\epsilon\biggl[\biggl(\dfrac{\sigma}{r_{ij}}\biggr)^{12}-\biggl(\dfrac{\sigma}{r_{ij}}\biggr)^6\biggr] + \phi_c,& \text{if } r_{ij}\leq r_c\\
		0,              & \text{if } r_{ij}> r_c
	\end{cases} 
\end{equation}
with $r_{ij}$, $\sigma$, $\epsilon$ the particles' distance, particles' diameter and potential well, respectively. The interaction radius $r_c$ is $2^{1/6}\sigma$ and the shift $\phi_c$ is the constant required to guarantee the continuity of the function in $r_c$.
The dissipation function associated to the system and the resulting TTCF is then 
\begin{equation}
	\label{eqn:TTCF1}
	\Omega=\dfrac{\dot{\gamma}V}{k_BT}P_{xy}\;\;\;\;\;\;\;\;\langle B(t)\rangle=\langle B(0)\rangle+\dfrac{\dot{\gamma}V}{k_BT}\int_0^t\langle P_{xy}(0)B(s)\rangle\text{d}s
\end{equation} 
with $V$ the volume of the system, $\dot{\gamma}$ the shear rate and $P_{xy}$ the shear pressure. 
The simulations  were carried out at the Lennard-Jones triple point ($\rho^*$ = $\rho\sigma^3$ = 0.8442; $T^*$ = $k_BT/\epsilon$ = 0.722). We varied the reduced strain  rate, $\gamma^*$ = $\gamma(m/\epsilon)^{1/2}\sigma$ from 1 to 10$^{-7}$. The shear viscosity $\eta$ was computed with the following TTCF expression
\begin{equation}
	\label{eqn:visc}
	\langle\eta(t)\rangle=\dfrac{\langle P_{xy}(t)\rangle}{\dot{\gamma}}=\dfrac{V}{k_BT}\int_0^t\langle P_{xy}(0)P_{xy}(s)\rangle\text{d}s
\end{equation} 
The timestep ($\Delta t^*$ = $\Delta t(m/\epsilon)^{1/2}/\sigma$) was 0.0025.
Each transient NEMD segment was 600 timesteps long, and we generated 4 $\times$ 30000 = 120000 of them, to yield reasonable statistics, with 30000 independent initial conditions and the corresponding mappings as described previously. The delay between samples along the equilibrium trajectory was 1000 timesteps.
We have used reduced units throughout, unless stated otherwise.
This first example is designed to be run on a local machine. As a result, the system is very small, and arguably the simplest system from which a realistic fluid response can be obtained. The total simulation time is approximately $4.5$ hours on a single core run for each shear rate tested. Since the method is fully parallelized, it is expected that a four-core run would be completed in approximately an hour. 

It is crucial that at the start of the daughter trajectory the correct velocity profile is superimposed to the momenta. This operation is required for the correct transient response, but is not automatically implemented in the LAMMPS routines for SLLOD dynamics. The operation can be performed with the command \emph{set atom}. \emph{velocity ramp} must be avoided for TTCF computation, since it overwrites the existing momenta with a linear profile, rather than superimposing it.

\begin{figure}
	\centering
	\includegraphics[width=\linewidth]{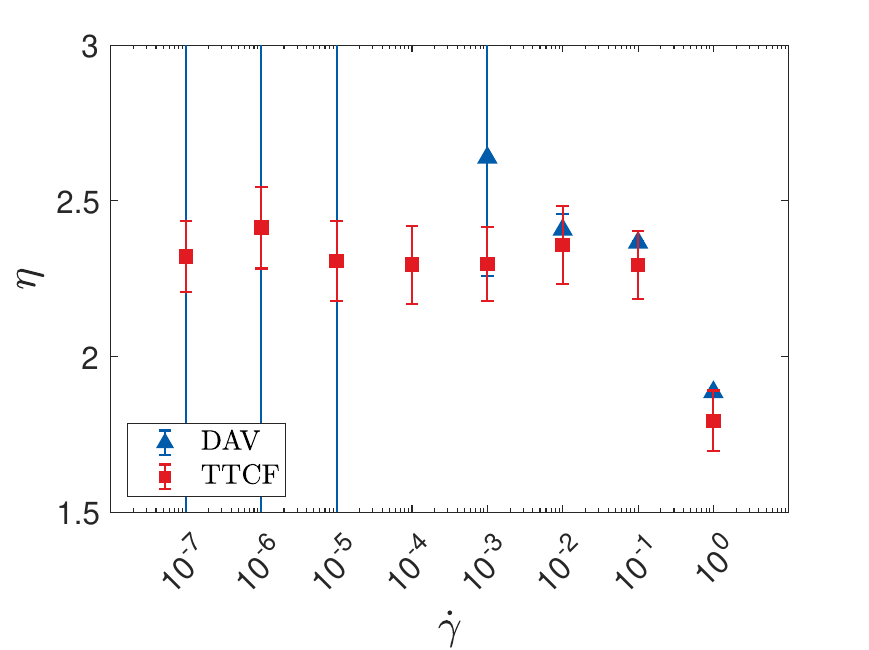}
	\caption{Shear viscosity for the bulk system at different shear rates ($5\times 10^4$-$5\times 10^11$ in MKS units). Both DAV and TTCF signals can detect the reduction in the viscosity for high shear rates (shear thinning). The accuracy of the direct average rapidly decreases for $\dot{\gamma}\to 0$. In contrast, the TTCF signal retains the same accuracy regardless of the magnitude of the driving force.}
	\label{fig:viscosity}
\end{figure}

Figure \ref{fig:viscosity} displays the shear viscosity computed via direct average (DAV) and TTCF. The data are in good agreement at high shear rates, where shear thinning occurs. As previously anticipated, the computed viscosity shows the feature of TTCF formalism of generating data whose accuracy is not affected by the magnitude of the external force. In contrast, the direct average (DAV) rapidly lose precision as the shear rate decreases. The effect will be explored more extensively in the next section.
We are aware that the current stable LAMMPS release (August 2023) contains several issues in the implementations of the SLLOD dynamics. These bugs are noticeable in the small discrepancy between the DAV and TTCF viscosities displayed, where the TTCF calculation is systematically larger than it direct counterpart. These issues are currently being analysed and addressed by Prof Debra Bernhardt (Searles) and her group\footnote{private communications.}. 

\section*{Example 2: shear of confined systems}

The confined system is based on that used in a previous study by Maffioli et al. \cite{Maffioli2022}, where a narrow three-dimensional channel was analyzed. The system is composed of Lennard-Jones particles, and the walls are kept in position by tethering the particles to lattice sites via harmonic springs. The boundary-driven Couette flow is generated by moving the lattices in opposite directions at constant velocity $v$ \cite{Ashurst1975}. 
The LJ potential is defined as 
\begin{equation}
	\phi(r)_{ij}= 
	4\epsilon\biggl[\biggl(\dfrac{\sigma}{r_{ij}}\biggr)^{12}-c_{ij}\biggl(\dfrac{\sigma}{r_{ij}}\biggr)^6\biggr]
\end{equation}
\noindent where the wetting parameter $c_{ij}$ is modulated to promote slip between wall and fluid particles, and to enhance cohesion within the fluid.
The equations of motion describing the dynamics of the system are:

\begin{equation}
	\label{eqn:system2}
	\begin{split}
		\dot{\textbf{r}}^f_i&=\dfrac{\textbf{p}^f_i}{m_i}\\
		\dot{\textbf{p}}^f_i&=-\sum_j\nabla\phi_{ij}\\
		\dot{\textbf{r}}^w_i&=\dfrac{\textbf{p}^w_i}{m_i}\\
		\dot{\textbf{p}}^w_i&=-\sum_j\nabla\phi_{ij}-k(\textbf{r}^w_i-\textbf{r}^l_i)-\alpha\textbf{p}^w_i\\
		\dot{\alpha}&=\dfrac{1}{Q}\biggl(\sum_i\textbf{p}^{w2}_i-3N^wk_BT\biggr)\\
		\dot{\textbf{r}}^l_i&=\bigl(\pm v\;\;,\;\;0\;\;,\;\;0\bigr)\\
		\dot{\textbf{p}}^l_i&=\bigl(0\;\;,\;\;0\;\;,\;\;0\bigr)
	\end{split}
\end{equation} 

\noindent with superscript $f$, $w$ and $l$ denoting the fluid, wall, and lattice particles respectively. $\textbf{r}$ and $\textbf{p}$ are the positions and the momenta of the particles, $k$ is the stiffness of the harmonic spring tethering the wall particles to the lattice sites. $\alpha$ is the Nos{\'{e}}-Hoover thermostat multiplier, acting on $N^w$ wall particles, and $T$ the target temperature of the walls.
The associated dissipation function is 
\begin{equation}
	\label{eqn:TTCF2}
	\Omega=\sum_ik(\textbf{r}^w_i-\textbf{r}^l_i)v
\end{equation}

\noindent summed over both walls.
In our previous work \cite{Maffioli2022}, we showed that it is also valid for inhomogeneous systems kept at constant pressure by means of a barostat \cite{Gattinoni2014}.
The wall temperature is set to $1$.
The system is composed of 6800 fluid particles and 2600 wall particles with diameter $\sigma=1$ and interaction radius $r_c=2.5$, at the densities of $0.7$ and $0.8$ respectively in reduced units.
The wetting parameter is $0.6325$ for the wall-fluid interactions and $1.2$ for the fluid-fluid potential \cite{thompson1990shear}.
The system is approximately $30\sigma$ long in the $x$ and $z$ direction, and the channel width is set to $10\sigma$.
This is expected to be sufficiently large in the lateral dimensions to avoid size effects on the liquid-solid friction \cite{Ogawa2019}.
The equations of motion were integrated with the standard velocity-Verlet \cite{Verlet1967} algorithm with integration step equal to $0.005$.

We adopted the first scheme described in the previous section, and $10^4$ independent mother trajectories were generated.
Each of them was thermalized for $5000$ time units before the sampling took place. The starting points were produced with a lag of $5$ time units from each other. $10^2$ initial states were generated for each mother, for a total of $4\times 10^6$ nonequilibrium daughter trajectories.
Each nonequilibrium system was monitored for 12.5 time units, or $2500$ time steps. 
Hence, a total of $2.1\times 10^{10}$ timesteps were required for each simulation.

The simulations were performed on the Swinburne University supercomputer OzSTAR, which features Intel Gold 6140 processors.
Each run, composed of $2.1\times 10^{6}$ time steps, was performed in approximately $6$ hours.
Since the resources needed for each single run were very limited, the total simulation was performed in a very short time.
If a higher efficiency is needed, each mother trajectory can be thermalized for shorter than $5000$ time units, which is arguably very conservative.
The mother trajectory was generated via the usual $fix\textunderscore nvt$ command. The same dynamics was applied on the daughter trajectories. The only difference in the dynamics between equilibrium and nonequilibrium runs was the constant velocity in the $x$ direction imposed to the lattice sites in the second case. As previously described, \emph{write\textunderscore restart} and \emph{read\textunderscore restart} commands were used to store and load the initial states. This also allowed us to run the mother trajectories one time only, and use the same sample for all the shear rates we tested.  
We computed the shear pressure $P_{xy}$ at the wall-fluid interface, using the method of planes \cite{Todd1995}, and the velocity profile.
We also compared the calculation of the friction coefficient using equilibrium methods with the direct measurement over the NEMD trajectories.
The equilibrium method is described in previous works \cite{hansen2011prediction,varghese2021improved}, and is based on the following definition:
\begin{equation}
	\langle F_x\rangle=\xi_0A\langle\Delta v \rangle
	\label{eqn:friction_coeff}
\end{equation} 
where $\xi_0$ is the intrinsic friction coefficient and $A$ the interface surface. $\langle F_x\rangle$ and $\langle\Delta v \rangle$ are respectively the instantaneous average force between the wall and the fluid particles located in a slab adjacent to the wall, and the instantaneous average slip velocity of the slab.
The data for the calculation were obtained from the window of $500$ time units along the mother trajectory, over which the starting points for the NEMD segments were generated. The DAV and TTCF data were instead produced by a direct nonequilibrium measurement of $\langle F_x\rangle$ and $\langle\Delta v \rangle$ from the daughter trajectories.
%
%
We investigated the fluid response to a shear rate spanning over 15 order of magnitude, from $10^{-1}$ to $10^{-15}$ in reduced units ($5\times10^{10}$\;-\;$5\times10^{-4}$ in SI units) .
This range is possibly the full window over which the system can be examined by numerical simulations.
Higher shear rates trigger a resonance with the harmonic bonds tethering the wall particles, resulting in wide wall oscillations and making the setup inefficient.
For weaker fields, the response might become hard to detect due to the finite precision of the floating-point representation, which is approximately $10^{-15}$ for double precision numbers. 

\begin{figure}
	\centering
	\includegraphics[width=\linewidth]{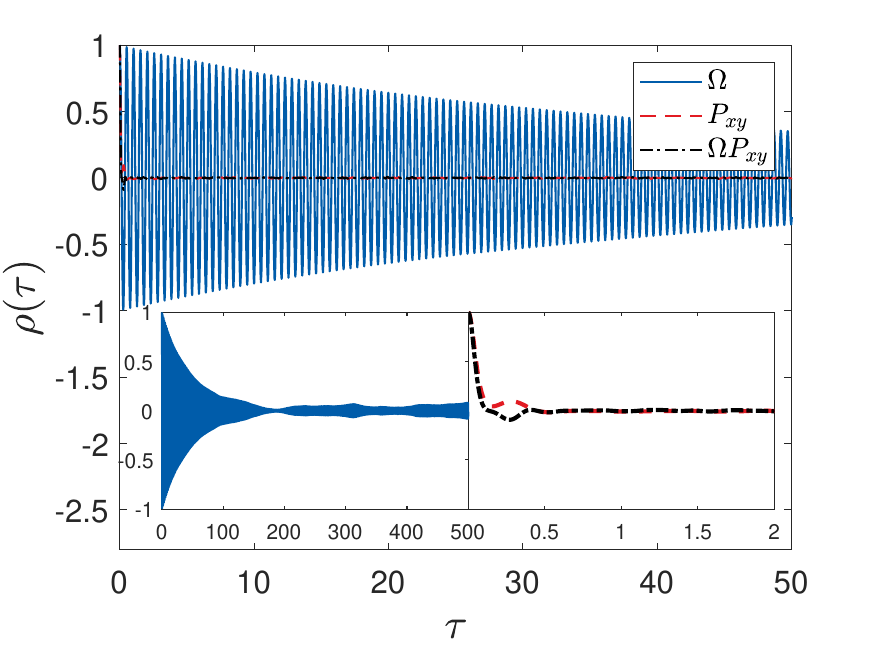}
	\caption{Autocorrelation of the dissipation function $\Omega$, the shear pressure $P_{xy}$ and the integrand function $\Omega P_{xy}$. Bottom left is the autocorrelation of $\Omega$ for a maximum lag of $500$ time units. Bottom right, the autocorrelation of $P_{xy}$ and $\Omega P_{xy}$, magnified with a narrower band of $2$ time units.}
	\label{fig:AutoCorr}
\end{figure}

In figure \ref{fig:AutoCorr}, the autocorrelation function for various quantities is displayed. The variables have been computed from a single equilibrium trajectory. 
Since at equilibrium the dissipation function is identically null, only the total wall-lattice harmonic force is displayed as $\Omega$ (cf. Eq. \ref{eqn:TTCF2}).
The presence of the harmonic springs combined with a weak wall-fluid interaction makes the wall effectively a solid and hence the autocorrelation function of $\Omega$ persists for a long time, in the order of $200$ time units, but does not fully decay.
On the other hand, $P_{xy}$ and $\Omega P_{xy}$ decorrelate in less than $1$ time unit, meaning that the starting points are reasonably independent for $T_s\ge 1$.
If the samples are correlated, the resulting signal might be both biased and more dispersed.
The short range decay in the autocorrelation is also possibly an indicator for the success of the TTCF method as it prevents the fluid from being in, or close to, the solid state, as noted earlier.
\label{eqn:mappings}

\begin{figure}
	\centering
	\begin{subfigure}{0.8\textwidth}
		\centering
		\includegraphics[width=1\linewidth]{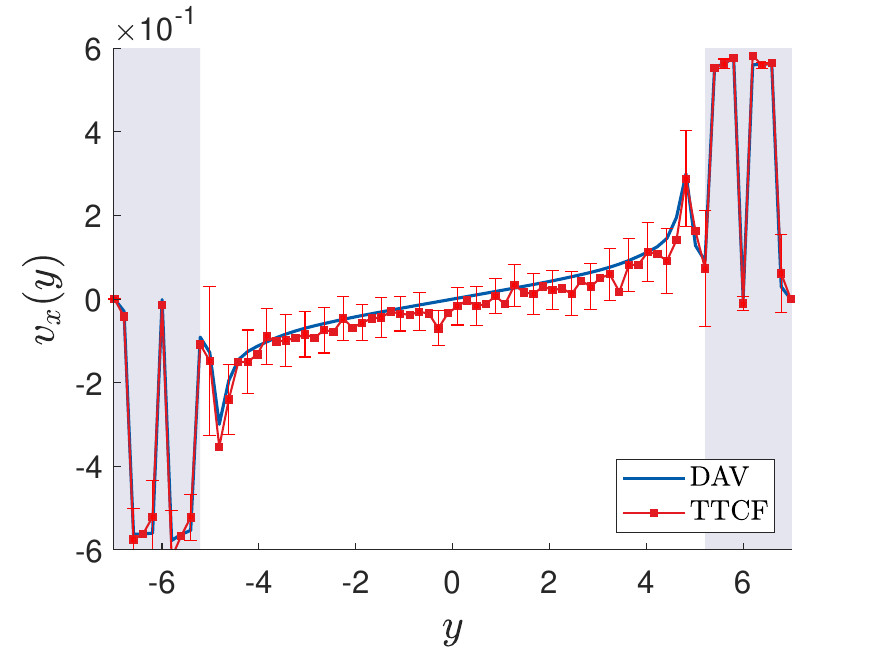}
		\caption{}
		\label{fig:vel_highshear}
	\end{subfigure}%
	\\
	\begin{subfigure}{0.8\textwidth}
		\centering
		\includegraphics[width=1\linewidth]{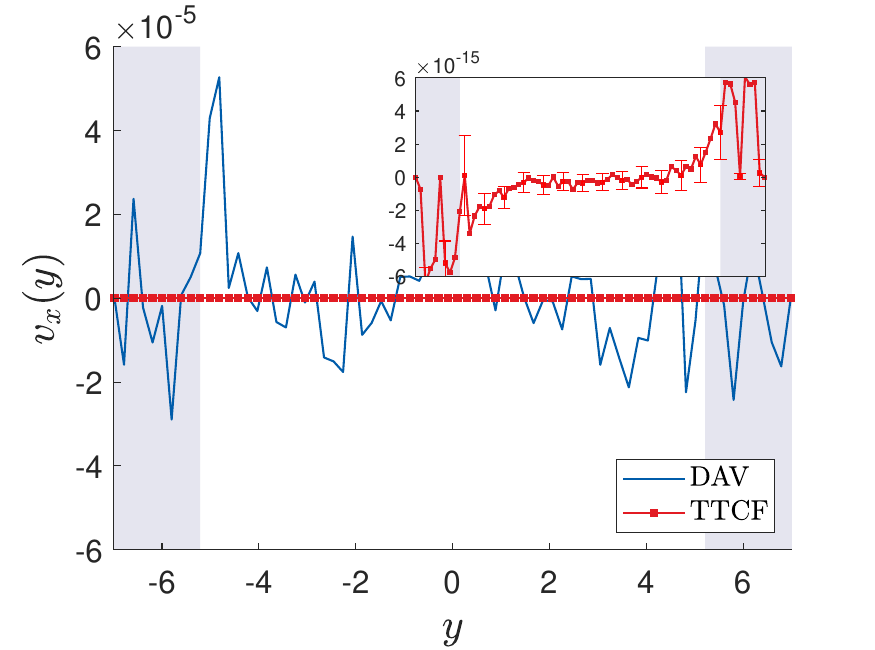}
		\caption{}
		\label{fig:vel_lowshear}
	\end{subfigure}
	\caption{Velocity profile at $t=12.5$ for $\dot{\gamma}=10^{-1}$ \textbf{(a)}, and $\dot{\gamma}=10^{-15}$ \textbf{(b)} (resulting wall velocity $v\approx\pm 10^2$ m/s and $v\approx\pm 10^{-12}$ m/s). The error bars are four times the standard error. The shaded regions indicate the solid walls.}
	\label{fig:Vel_profile}
\end{figure}

\begin{figure}
	\centering
	\begin{subfigure}{0.8\textwidth}
		\centering
		\includegraphics[width=1\linewidth]{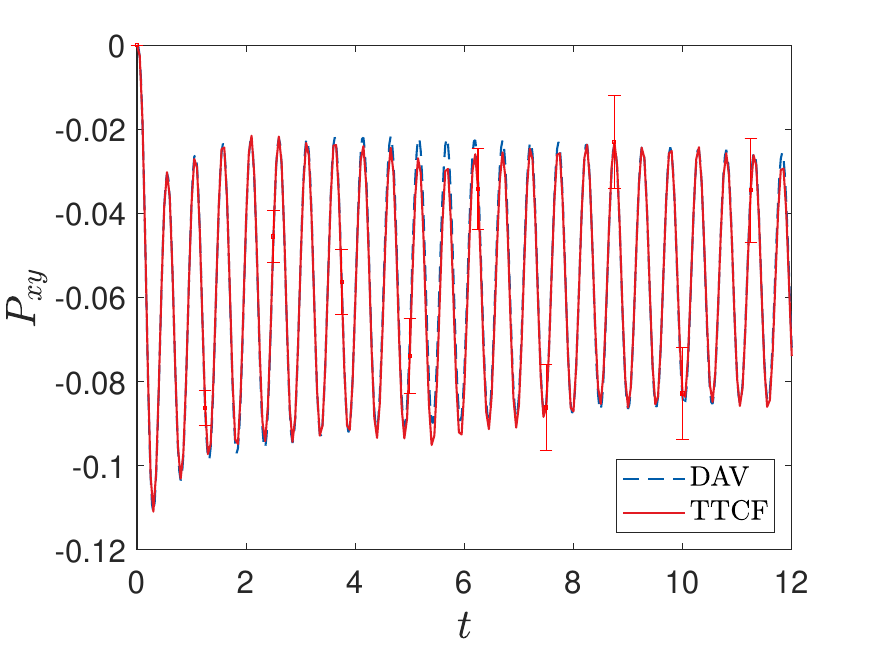}
		\caption{}
		\label{fig:Pxy_highshear}
	\end{subfigure}%
	\\
	\begin{subfigure}{0.8\textwidth}
		\centering
		\includegraphics[width=1\linewidth]{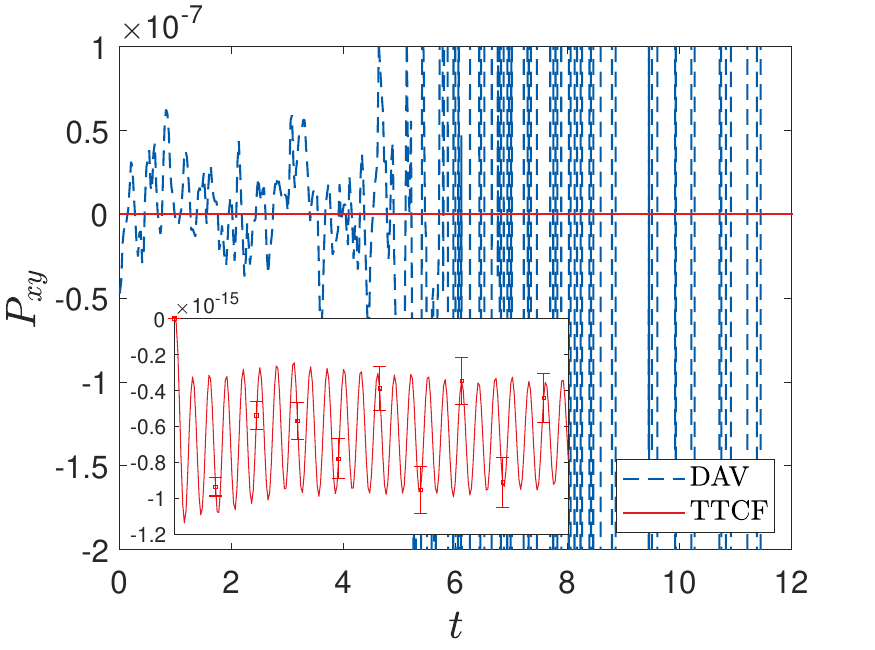}
		\caption{}
		\label{fig:Pxy_lowshear}
	\end{subfigure}
	\caption{Shear pressure as a function of time for $\dot{\gamma}=10^{-1}$ \textbf{(a)}, and $\dot{\gamma}=10^{-15}$ \textbf{(b)} ($5\times10^{10}\text{s}^{-1}$ and $5\times10^{-4}\text{s}^{-1}$ in SI units). The error bars are four times the standard error.}
	\label{fig:Pressure}
\end{figure}

\begin{figure}
	\centering
	
	\includegraphics[width=\linewidth]{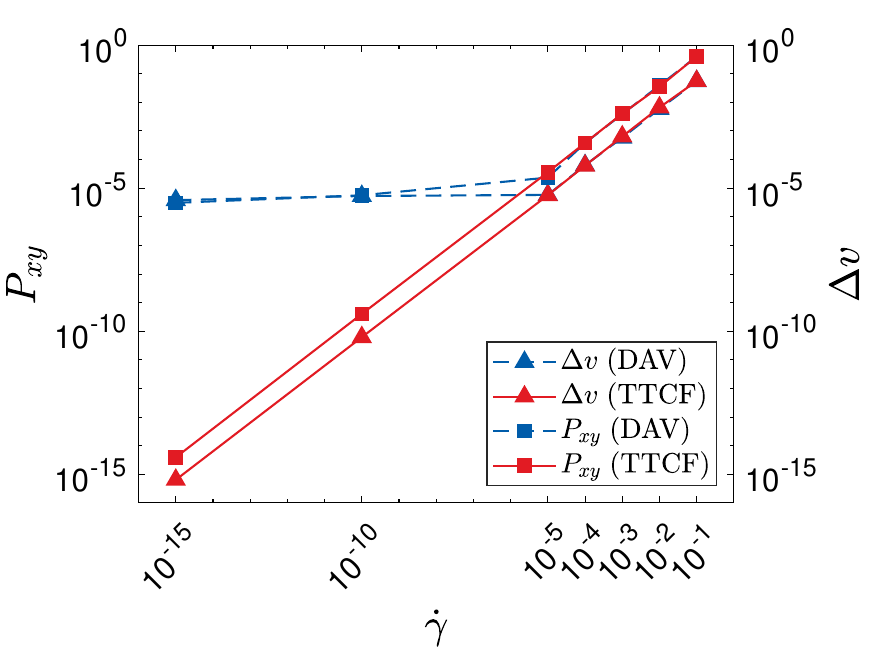}
	
	\caption{Shear pressure and velocity profile for DAV and TTCF for every shear rate tested spanning from $5\times10^{10}\text{s}^{-1}$ to $5\times10^{-4}\text{s}^{-1}$ in SI units. Error bars have been omitted.}
	\label{fig:Pressure_slip}
\end{figure} 

Figure \ref{fig:Vel_profile} shows the velocity profile at $t=12.5$ for $\dot{\gamma}=10^{-1}$ and $10^{-15}$ computed via DAV and  TTCF.
The error bars for the direct average have been omitted since the signal is either extremely accurate  for the high shear, or extremely noisy for the weak field.
The sigmoid shape of the velocity profile is the result of the small channel width combined with the Lennard-Jones interaction parameters chosen, and it persists in the steady state.
In Figure \ref{fig:Pressure} the transient of the shear pressure at the wall-fluid interface is displayed from $t=0$.
The error bars for the direct average have been omitted again. 
Both the velocity and the pressure show that for $\dot{\gamma}=10^{-15}$ the DAV signal is essentially composed of random noise. 
In Figure \ref{fig:Pressure_slip} the shear pressure and the slip velocity as a function of the shear rate are displayed.
%
%
The slip velocity is defined as the difference of the velocity of the innermost wall layer and the outermost fluid layer.
The data show that the DAV can produce acceptable results for $\dot{\gamma}\ge 10^{-5}$. For weaker fields, the statistical fluctuations drown out the response, and the signal is constant across the different shear rates.

On the other hand, all the phase variables displayed suggest that the TTCF method is able to identify the response at any level of the shear rate. The accuracy of the TTCF method increases as the shear rate decreases and the signal-to-noise ratio (SNR), defined as the ratio between the signal and its standard deviation, remains approximately constant for any magnitude of the shearing force.
The property can be proven as follows.
For simplicity, the fluctuations of any quantity (e.g. $P_{xy}$) are assumed to be constant for every shear rate. 
This condition is valid for weak shear rates, and it essentially holds across the entire range tested in this work.
Secondly, the dissipation function at $t=0$ is proportional to the shear rate. This relation is exact and independent of the rate itself, and it immediately follows from Eq. (\ref{eqn:TTCF1}) and (\ref{eqn:TTCF2}), where in the expression of the dissipation function two terms are clearly detectable. The first term is a pressure/force contribution ($P_{xy}$ for the bulk systems, $x$ component of the spring forces for the channel), and in statistical mechanics is usually denoted as dissipative flux conjugate to the generalized external field. For our purpose, we note that this is computed at $t=0$, that is, an \textit{equilibrium} state, and hence is independent of the external field. The second term is the effective external force and is equal, or proportional, to $\dot{\gamma}$ or $v$. 
As a result, $\Omega(0)$ is proportional to the magnitude of the external driving force, and we have the following set of relations, shown for simplicity only for the shear pressure and the dissipation function:

\begin{equation}
	\begin{split}
		\langle\Omega(0)\rangle=0 \; \; \; \; \; \sigma(\Omega(0))=k_1\dot{\gamma}\\
		\langle P_{xy}(t)\rangle=\hat{P}_{xy}(\dot{\gamma}) \; \; \; \; \; \sigma(P_{xy}(t))=k_2\\
	\end{split}
\end{equation}

\noindent where $\sigma$ is the standard deviation, $k_i$ are arbitrary constants, and $\hat{P}_{xy}$ is the real shear pressure.
If we restrict, for convenience, our analysis to a steady state, the functions $\Omega(0)$ and $P_{xy}$ are uncorrelated and the variance $\sigma^2$ of the integrand function in Eq. (\ref{eqn:TTCF1}) and (\ref{eqn:TTCF2}) is:

\begin{equation}
	\begin{split}
		\sigma^2(\Omega(0)P_{xy}(t))=&\\
		&(\sigma^2(\Omega(0))+\langle\Omega(0)\rangle^2)(\sigma^2(P_{xy}(t))\\
		+&\langle P_{xy}(t)\rangle^2)-\langle\Omega(0)\rangle\langle P_{xy}(t)\rangle\\
		=&(k_1^2\dot{\gamma}^2)(k_2^2+\hat{P}_{xy}^2)
	\end{split}
	\label{eqn:variance}
\end{equation}

\noindent The effect of the integration can hardly be modeled by simple functions, but it depends solely on the time $t$, and hence the SNR of the shear pressure 
\begin{equation*}
	\mathrm{SNR}=\dfrac{\langle P_{xy}(t)\rangle}{\sigma(P_{xy}(t))}
\end{equation*}

\noindent is expressed by the following identities:

\begin{equation}
	\begin{split}
		\mathrm{SNR}_{\mathrm{DAV}}&=\dfrac{\hat{P}_{xy}(\dot{\gamma})}{k_2}\\
		\mathrm{SNR}_{\mathrm{TTCF}}&=K(t)\dfrac{\hat{P}_{xy}(\dot{\gamma})}{k_1\dot{\gamma}\sqrt{k^2_2+\hat{P}_{xy}^2(\dot{\gamma})}}=\dfrac{K(t)}{k_1\dot{\gamma}}\dfrac{\mathrm{SNR}_{\mathrm{DAV}}}{\sqrt{1+\mathrm{SNR}^2_{\mathrm{DAV}}}}   
	\end{split}
	\label{eqn:SNR}
\end{equation}

\noindent where $K(t)$ is a function accounting for the integration process. Since $\hat{P}_{xy} \simeq c \dot{\gamma}$ for $\dot{\gamma} \to 0$, we have
\begin{equation}
	\begin{split}
		&\lim_{\dot{\gamma}\to 0}\mathrm{SNR}_{\mathrm{DAV}}=0\\
		&\lim_{\dot{\gamma}\to 0}\mathrm{SNR}_{\mathrm{TTCF}}=\dfrac{K(t)c}{k_1}\\  
	\end{split}
	\label{eqn:SNR0}
\end{equation}
and
\begin{equation}
	\begin{split}
		&\lim_{\dot{\gamma}\to \infty}\mathrm{SNR}_{\mathrm{DAV}}=\infty\\
		&\lim_{\dot{\gamma}\to \infty}\mathrm{SNR}_{\mathrm{TTCF}}=0.\\  
	\end{split}
	\label{eqn:SNRinf}
\end{equation}
The last relations suggest that the TTCF methods are eventually outperformed by a simple direct average, and that they progressively lose accuracy as the external field is increased. 

\begin{figure}
	\centering
	\includegraphics[width=\linewidth]{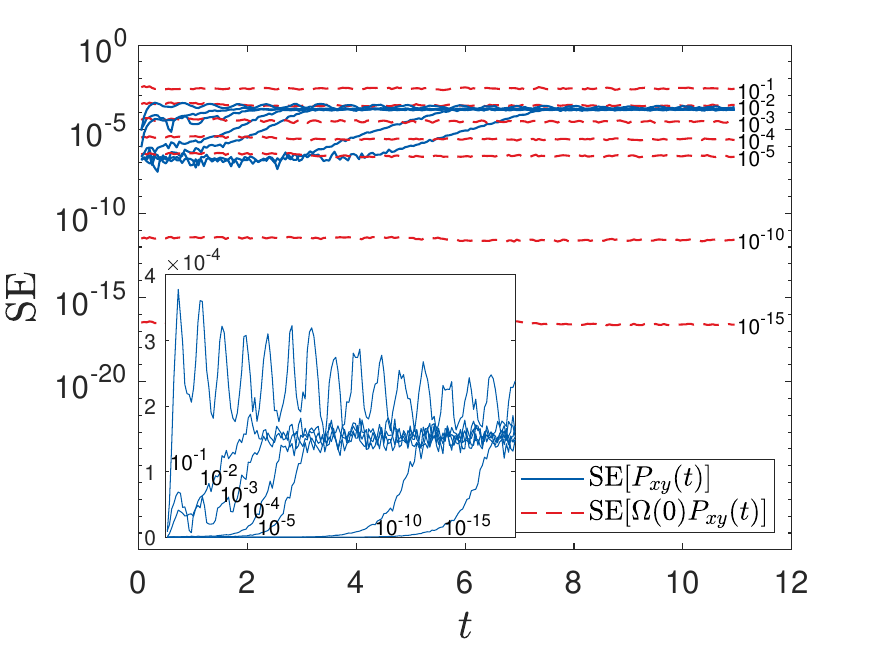}
	\caption{Standard error of $P_{xy}(t)$ and of $\Omega(0) P_{xy}(t)$. The inset shows the first quantity only, magnified. The different curves are for different shear rates.}
	\label{fig:SE}
\end{figure}

Figure \ref{fig:SE} summarizes the various results: the standard error SE of $P_{xy}(t)$ and $\Omega(0)P_{xy}(t)$ are displayed for all the shear rates ($\mathrm{SE}=\sigma/\sqrt{N}$ with $N=4\times 10^{6}$ the total number of daughter trajectories). The latter is approximately constant along the NEMD segment, and proportionally increases with $\dot{\gamma}$. In the steady states, the DAV fluctuations are independent of $\dot{\gamma}$, while they exponentially increase from a null value during the transient. This effect occurs because the mappings selected guarantee that $\langle \Omega(0) \rangle=0$, but also $\langle P_{xy}(0) \rangle=0$. In other words, the averaging over the mappings artificially eliminate the instantaneous fluctuations of the pressure, and the accuracy of the direct average at $t=0$ is virtually infinite. This correlation is retained for the first moments of the NEMD trajectory, whose signal is still characterized by low statistical uncertainty. The chaotic nature of the system makes the trajectories exponentially diverge, and, as a result, the standard error increases until the trajectories become uncorrelated, and the fluctuations stabilize to a steady value. After this, a measure obtained from a sample of $N$ initial states along with their three mappings is equivalent to a set of $4N$ independent states.
Similar features are exploited in previous works \cite{Ciccotti1979}, where, in order to improve the accuracy of traditional NEMD simulations, the nonequilibrium segments are paired with equilibrium ones starting from the same initial condition. The quantity of interest is monitored over both trajectories and the final result is the difference between the two measures. At the start of the run, where the two trajectories are highly correlated, the procedure eliminates from the NEMD signal the random fluctuations, obtained from the equilibrium signal. However, the systems studied in NEMD simulation are almost invariably chaotic, hence the trajectories decorrelate rapidly, and the obtained signal may lose most of its accuracy well before the system has reached a steady state.
The rate of divergence of the trajectories is also driven by the magnitude of the shear: strong fields make the system more chaotic. This so-called \textit{background subtraction} technique was demonstrated to be unsutable for practical systems of interest \cite{evans1990statistical}.

\begin{figure}
	\centering
	
	\includegraphics[width=\linewidth]{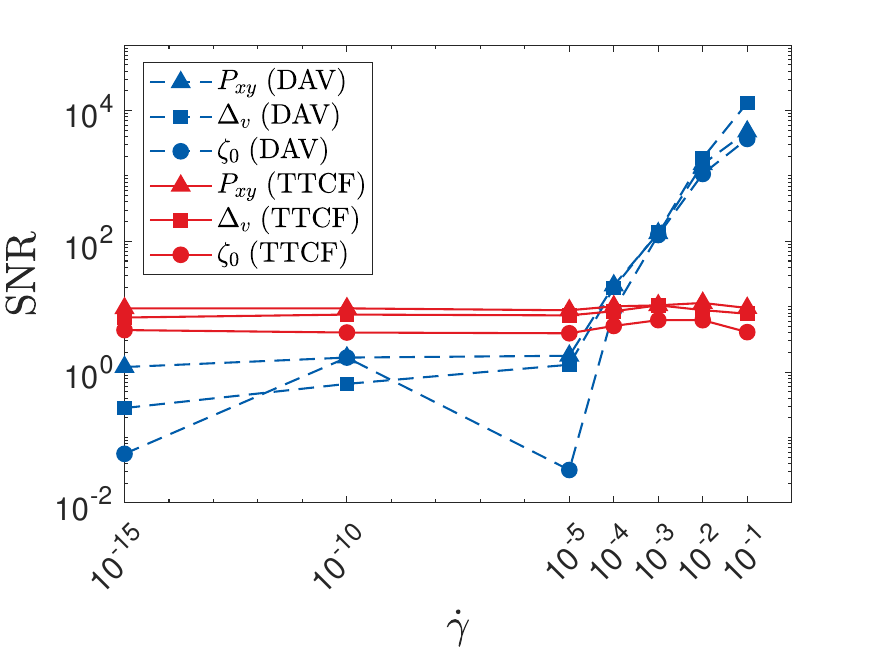}
	
	\caption{Signal-to-noise ratio for various quantities for DAV and TTCF as a function of the reduced shear rate. }
	\label{fig:SNR}
\end{figure}

In Figure \ref{fig:SNR} the signal-to-noise ratio is displayed, for the shear pressure, the slip velocity and the friction coefficient for both DAV and TTCF.
The threshold below which the TTCF outperforms the direct average is confirmed to be $\dot{\gamma}=10^{-5}$ for all the quantities.
A SNR of the order of $1$ or below indicates that the magnitude of the statistical fluctuations are comparable or larger than the signal, and no information can be obtained from the data.

\begin{figure}
	\centering
	\includegraphics[width=\linewidth]{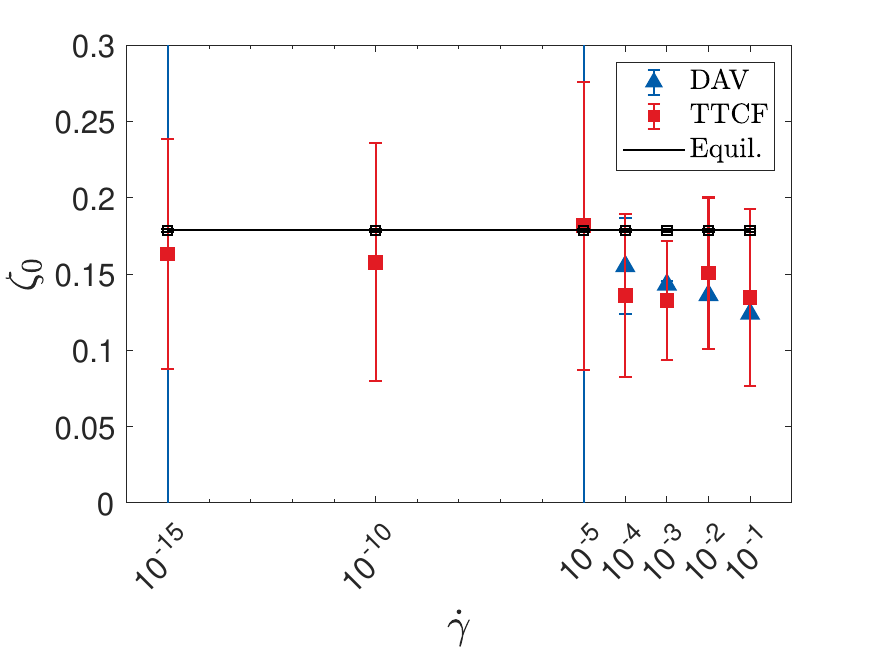}
	\caption{Friction coefficient as a function of the shear rate, computed with DAV and TTCF. The black squares indicate the (constant) friction coefficient computed using the method proposed in Refs\cite{hansen2011prediction,varghese2021improved}. The DAV signal   
		at $\dot{\gamma}=10^{-5}$ and $\dot{\gamma}=10^{-15}$ does not appear in the range of the plot, and for $\dot{\gamma}=10^{-10}$ the confidence interval itself is out of scale.}
	\label{fig:friction_coeff}
\end{figure}

Figure \ref{fig:friction_coeff} displays the friction coefficient, computed using the equilibrium and nonequilibrium methods. The black line represents the data computed via the equilibrium method. Since the equilibrium calculation is not associated to a specific shear rate, the line is constant across the figure, and displayed for the mere purpose of comparison with the NEMD measurements.
The data show the same behaviour as previous quantities, with DAV being progressively less accurate and the TTCF retaining the same precision. 
In comparison, the equilibrium method is drastically better at any level.
For shear rates beyond $10^5$ the accuracy of the DAV is comparable to the equilibrium method, and the friction coefficient starts to deviate from the linear response.

The friction coefficient is particularly hard to estimate via direct average because a ratio of two random variables can be highly unstable. The effect is severe at low shear rate, where the signal may even be out of the numeric range displayed here.

The results confirm the model for the SNR previously derived, with the caveat that the numerator of the direct average must be replaced by the real response, computed, for instance, with the TTCF method.
The TTCF signal-to-noise ratio is approximately constant for every shear rate, and it is expected to decrease for higher shear rates, when deviations from linear response become relevant. The effect might already starts at $\dot{\gamma}=10^{-1}$, where the SNR slightly decreases. We note that for more complex molecular fluids, such as alkanes or polymer melts in solutions, nonlinear effects manifest at much weaker shear rates, making TTCF the technique of choice for its superior statistical accuracy at experimental strain rates.

\section*{Conclusion}

We applied the TTCF technique in the investigation of an atomic fluid confined in a narrow channel undergoing planar Couette flow.
We have shown that the method is readily implementable in simulation software such as LAMMPS and can be generalized to molecular fluids, and we listed the key ingredients for its efficient usage.
We highlighted the advantages of the TTCF formalism, particularly the property of retaining the same signal-to-noise ratio for arbitrarily weak fields, to the point that the lower bound is now set by the finite precision of the computer arithmetic.
We provided a simple proof of this phenomenon and, given the generality of the mechanism under which it occurs, we expect it to hold for other types of driving fields.   

The application of the TTCF formalism demands significant computational resources, but the relative accuracy of the data produced is independent of the magnitude of the external driving field, and it rapidly outperforms the direct average for weak driving fields, which correspond to those that can be experimentally probed in the laboratory.
As a comparison, the resources needed to monitor the system response at $\dot{\gamma}=10^{-15}$ via DAV could be as much as $10^{20}$ times larger than those needed for the TTCF algorithm.
Additionally, we have shown that the simulations can be split into a massive number of short, independent tasks, hence dramatically increasing their efficiency.

\section*{Acknowledgements}

We thank the Australian Research Council for a grant obtained through the Discovery Projects Scheme (Grant No. DP200100422) and the Royal Society for support via International Exchanges, Grant No. IES/R3/170/233.
J.P.E. was supported by the Royal Academy of Engineering (RAEng) through their Research Fellowships scheme.
D.D. was supported through a Shell/RAEng Research Chair in Complex Engineering Interfaces.
The authors acknowledge the Swinburne OzSTAR Supercomputing facility and the Imperial College London Research Computing Service (DOI:10.14469/hpc/223) for providing computational resources for this work.
We thank Debra Bernhardt and Stephen Sanderson (University of Queensland) for useful discussions regarding the implementation of SLLOD in LAMMPS.

\bibliographystyle{ieeetr} 
\bibliography{bibliography}

\end{document}